\documentclass[aps,prx,reprint,twocolumn,superscriptaddress,nofootinbib]{revtex4-2}
\usepackage{graphicx,amsmath,amsfonts,amssymb,xr}
\usepackage{epsfig, color, dsfont, upgreek, physics}
\usepackage{mathrsfs}
\usepackage{mathtools}
\usepackage{bbold}
\usepackage{comment}
\usepackage{braket}
\usepackage{cancel}
\usepackage{lipsum}
\usepackage{soul}
\usepackage{bm}
\usepackage{appendix}
\usepackage{ulem}
\usepackage{float}

\usepackage[bookmarks=true,colorlinks,linkcolor=OrangeRed,urlcolor=NavyBlue,citecolor=RoyalBlue]{hyperref}
\DeclareMathAlphabet{\mathbbold}{U}{bbold}{m}{n}
\usepackage[capitalise]{cleveref}
\Crefname{figure}{Fig.}{Figs.}
\Crefname{section}{Sec.}{Secs.}
\usepackage[dvipsnames]{xcolor}
\usepackage[inline]{enumitem}
\setlist[enumerate]{label=(\roman*)}
\usepackage{cmds}
\usepackage{stackengine, scalerel}

\graphicspath{{figures/}}
\usepackage{orcidlink}
\makeatletter
\AtBeginDocument{%
  \def\selectlanguage#1{}%
  \def\otherlanguage#1{}%
}
\makeatother

\definecolor{revcolor}{HTML}{8A0FBF}

\definecolor{rwcolor}{HTML}{0E7C7B}

\newcommand{\SRE}[1][2]{\mathcal{M}_{#1}}                     
\newcommand{\PE}{\mathrm{PE}}                                 

\footnotetext{jad.halimeh@lmu.de\\emanuele.tirrito@epfl.ch}

\newcommand{\LMU}{\affiliation{Department of Physics and Arnold Sommerfeld Center for Theoretical Physics (ASC), Ludwig Maximilian University of Munich, 80333 Munich, Germany}}
\newcommand{\MPQ}{\affiliation{Max Planck Institute of Quantum Optics, 85748 Garching, Germany}}
\newcommand{\MCQST}{\affiliation{Munich Center for Quantum Science and Technology (MCQST), 80799 Munich, Germany}}
\newcommand{\KYUNGHEE}{\affiliation{Department of Physics, College of Science, Kyung Hee University, Seoul 02447, Republic of Korea}}
\newcommand{\EPFL}{\affiliation{Institute of Physics, École Polytechnique Fédérale de Lausanne (EPFL), CH-1015 Lausanne, Switzerland}}
\newcommand{\CQSE}{\affiliation{Center for Quantum Science and Engineering, École Polytechnique Fédérale de Lausanne (EPFL), CH-1015 Lausanne, Switzerland}}

\begin{document}
\author{Lukas Ebner\,
\orcidlink{0009-0005-5186-9042}} \LMU \MPQ \MCQST
\author{Giovanni Cataldi\,
\orcidlink{0000-0002-9073-8978}} \MPQ \MCQST
\author{Kaidi Xu\,\orcidlink{0000-0003-2184-0829}} \LMU \MPQ \MCQST
\author{Jad C. Halimeh\,\orcidlink{0000-0002-0659-7990}$^\ast$}\LMU \MPQ \MCQST \KYUNGHEE
\author{Emanuele Tirrito
\,\orcidlink{0000-0001-7067-1203}$^\ast$} \EPFL \CQSE

\title{Global and nonlocal magic of quantum many-body scars}
\begin{abstract}
    Nonergodic features of chaotic quantum many-body systems are commonly characterized through local observables, fidelity, and entanglement entropy. Here, using global and nonlocal measures of nonstabilizerness (magic), we study quantum many-body scars near infinite temperature in $(1+1)$-dimensional Abelian $\mathbb{Z}_2$ and U$(1)$ lattice gauge theories, both analytically and numerically. We derive an exact closed-form expression for the nonlocal trace distance magic solely from the Schmidt spectrum, reducing the optimization over local unitaries to a finite maximization over stabilizer-compatible Schmidt ranks. We find that scar eigenstates can exhibit extensive global magic despite their anomalously low entanglement, while retaining anomalously large nonlocal magic compared with ergodic states. We further show that constrained Hilbert spaces can generate irreducible nonlocal magic through stabilizer-incompatible Schmidt ranks, even for a flat entanglement spectrum.
\end{abstract}

\maketitle
\section{Introduction}

Whether and how an isolated quantum many-body system reaches thermal equilibrium under its own unitary dynamics is a fundamental question in understanding the emergence of statistical thermodynamic behavior from microscopic quantum dynamics~\cite{Rigol2008ThermalizationItsMechanism,%
DAlessio2016QuantumChaosEigenstate}. 
One prominent answer is the eigenstate thermalization hypothesis (ETH), which posits that individual highly excited eigenstates of a generic non-integrable Hamiltonian already encode thermal expectation values, so that local observables relax to their microcanonical predictions and memory about the details of the initial state is erased~\cite{Deutsch1991QuantumStatisticalMechanics, 
Srednicki1994ChaosQuantumThermalization, Deutsch2018EigenstateThermalizationHypothesis}.

Over the past decade, a paradigm of weak ergodicity breaking has challenged this picture without recourse to disorder:
quantum many-body scars (QMBS)~\cite{Turner2018WeakErgodicityBreaking, Turner2018QuantumScarredEigenstates, Serbyn2021QuantumManybodyScars, Moudgalya2022QuantumManybodyScars}.
A scarred system hosts a vanishing fraction of atypical, roughly equally spaced eigenstates that sit at high temperature yet carry anomalously low, subthermal entanglement~\cite{Lin2019ExactQuantumManyBody, Shiraishi2017SystematicConstructionCounterexamples, Moudgalya2024ExhaustiveCharacterizationQuantum, Mark2020UnifiedStructureExact}, so that initial states with large overlap on this subspace evade thermalization and produce long-lived coherent revivals~\cite{Bernien2017ProbingManybodyDynamics, Iadecola2020QuantumManybodyScar, Moudgalya2024ExhaustiveCharacterizationQuantum}. 
Diagnosing scars, however, has relied predominantly on fidelity revivals, local observables and bipartite entanglement. The latter probes one specific notion of state complexity, where entanglement is the resource and separable states are the free states. Other notions of complexity arise from different resource theories, one of which has attracted considerable attention in recent years: nonstabilizerness, or magic. 

The stabilizer framework provides a classically efficient description of quantum states and dynamics: stabilizer states and Clifford circuits can be efficiently simulated on a classical computer, as formalized by the Gottesman--Knill theorem~\cite{Gottesman1997StabilizerCodesQuantum, Bravyi2016ImprovedClassicalSimulation, Bravyi2012MagicstateDistillationLow}. A measure of nonstabilizerness quantifies the resources to move beyond this stabilizer framework, providing a different notion of quantum state complexity than entanglement~\cite{Veitch2014ResourceTheoryStabilizer, Howard2017ApplicationResourceTheory}. 

Conceptually, entanglement and magic quantify distinct aspects of quantum state complexity---a state can be highly entangled yet stabilizer-like, or not entangled yet strongly magical; nevertheless, they do not seem to be completely independent~\cite{turkeshi2023,tirrito2024,frau2024,iannotti2025entanglement,gu2025}. The Stabilizer-Rényi entropy (SRE)~\cite{Leone2022StabilizerRényiEntropy, Leone2024StabilizerEntropiesAre, Liu2022ManyBodyQuantumMagic} has made magic quantitatively accessible in many-body systems~\cite{Oliviero2022MagicstateResourceTheory, White2021ConformalFieldTheories, Viscardi2026InterplayEntanglementStructures}, although its evaluation typically remains exponentially costly. The SRE measure of magic has since been applied to ground-state criticality~\cite{Oliviero2022MagicstateResourceTheory, White2021ConformalFieldTheories, Viscardi2026InterplayEntanglementStructures}, quantum chaos~\cite{Turkeshi2025PauliSpectrumNonstabilizerness, Turkeshi2025MagicSpreadingRandom}, and thermalizing dynamics~\cite{Tirrito2025AnticoncentrationNonstabilizernessSpreading, Odavic2025StabilizerEntropyNonintegrable, Tirrito2025UniversalSpreadingNonstabilizerness}. Since ergodic evolution drives a state toward random-state behavior with nearly maximal magic~\cite{Turkeshi2025PauliSpectrumNonstabilizerness, Kim2024AverageRényiEntropy}---magic is a natural diagnostic for nonergodicity: a scar, failing to thermalize, should fail to become typically magical. 
Yet nonstabilizerness remains a largely unexplored diagnostic of weak ergodicity breaking.

Lattice gauge theories (LGTs)~\cite{Wilson1974ConfinementQuarks, Kogut1975HamiltonianFormulationWilsons, Susskind1977LatticeFermions, Kogut1979IntroductionLatticeGauge} provide the natural arena in which these threads meet, and are today a central target of analog and digital quantum simulation~\cite{Martinez2016RealtimeDynamicsLattice, Yang2020ObservationGaugeInvariance,Zhou2022ThermalizationDynamicsGauge, Calajo2024DigitalQuantumSimulation,Gyawali2025ObservationDisorderfreeLocalization,Cochran2025VisualizingDynamicsCharges,GonzalezCuadra2025,cobos2025realtimedynamics21dgauge,xu2026observationglueballexcitationsstring,joshi2026observationgenuine21dstring,gandon2026stringdynamics21du1,ciavarella2026quantumsimulationstwodimensionalnonabelian}. Gauge invariance imposes local constraints, which are now understood to underlie some of the most prominent instances of QMBS, Hilbert-space fragmentation, and disorder-free localization~\cite{Surace2020LatticeGaugeTheories, Halimeh2023RobustQuantumManybody, Aramthottil2022ScarStatesDeconfined, Cataldi2025DisorderFreeLocalizationFragmentation}.
Indeed, quantum scarring occurs across a broad range of gauge theories: the paradigmatic PXP model realized in Rydberg arrays, which hosts prominent quantum scars~\cite{Bernien2017ProbingManybodyDynamics}, can be mapped onto a spin-$1/2$ $\mathrm{U}(1)$ Schwinger LGT~\cite{Surace2020LatticeGaugeTheories, Turner2018QuantumScarredEigenstates, Desaules2023ProminentQuantumManybody}; scarring also occurs in Abelian $\mathbb{Z}_2$ and $\mathrm{U}(1)$ models with exactly known scars~\cite{Iadecola2020QuantumManybodyScar}, and extends to non-Abelian $\mathrm{SU}(2)$ Yang--Mills theory with dynamical matter~\cite{Calajo2025QuantumManybodyScarring, Cataldi2024Simulating2+1DSU2}.
These same constraints reshape quantum resources: gauge invariance slows and bounds the spreading of magic~\cite{Falcao2025NonstabilizernessU1Lattice, Tarabunga2023ManyBodyMagicPauliMarkov, esposito2025magic,Smith2025NonstabilizernessKineticallyConstrained}. 
An interesting question is whether magic can serve as a diagnostic of scarring itself---and, if so, which notions of magic are most informative.

Here lies a subtlety that motivates our central distinction. 
The global SRE conflates resource content that is genuinely delocalized with magic that might be removable by unitary local basis rotations. To separate these contributions, one can define nonlocal magic---the nonstabilizerness that remains after minimizing over all local unitary basis changes within the considered subsystems. The resulting quantity is invariant under local basis changes. In this respect, it shares an important feature with the entanglement entropy, which is likewise insensitive to local changes of basis~\cite{Korbany2025LongRangeNonstabilizernessPhases, White2021ConformalFieldTheories, Viscardi2026InterplayEntanglementStructures}.

The flatness of the entanglement spectrum controls the nonlocal magic~\cite{Cao2025GravitationalBackreactionMagical}: for Haar-random pure states, the reduced density matrix (RDM) of a sufficiently small subsystem is typically close to maximally mixed, with an approximately flat spectrum~\cite{Page_1993}, suggesting that the nonlocal magic is small, whereas locally structured, non-generic states need not follow this behavior. This suggests that global and nonlocal magic may tell opposite stories about scars.
The recent discovery of scar states that are stabilizer states~\cite{Hartse2025StabilizerScars, Gupta2026ExactStabilizerScars}, and thus carry zero magic, further raises the question of how generic this property is among scar states.

In this work, we characterize QMBS, which lie close to infinite temperature, $\beta=1/T \approx 0$, through the lens of magic in two constrained models: the PXP/$\mathrm{U}(1)$ Rydberg-blockade chain~\cite{Turner2018WeakErgodicityBreaking, Surace2020LatticeGaugeTheories} and a $\mathbb{Z}_2$ model. Both host exactly solvable scar states~\cite{Iadecola2020QuantumManybodyScar, Halimeh2023RobustQuantumManybody}.
Using the SRE to quantify the global magic~\cite{Leone2022StabilizerRényiEntropy, Liu2025QuantumAlgorithmsInverse, Tirrito2025UniversalSpreadingNonstabilizerness}, together with an exact expression for the nonlocal trace-distance magic and an SRE-based upper bound on nonlocal magic~\cite{Viscardi2026InterplayEntanglementStructures, Korbany2025LongRangeNonstabilizernessPhases, Kim2024AverageRényiEntropy}, we uncover qualitative differences between the resource structure of QMBS and that of standard ergodic states.

\section{Quantum complexity measures}
We characterize the computational complexity of the states studied below through two complementary notions of magic. 
The global magic quantifies the total nonstabilizerness of a state. This quantity is closely related to the cost of simulating it classically~\cite{Bravyi_2005,Bravyi2012MagicstateDistillationLow,Howard2017ApplicationResourceTheory}. The global magic is Clifford-invariant but not invariant under general local unitaries.
The nonlocal magic retains only the part of that resource that cannot be removed by local unitary basis changes applied to the two subsystems, and is therefore a property of the entanglement structure rather than of the local basis~\cite{Cao2025GravitationalBackreactionMagical,iannotti2025entanglement,collura2026nonlocal,iannotti2026non,sierant2026exact}. 
The two notions ask different questions of a scarred state, and, as we show below, they receive different answers.

Throughout, $\ket{\psi}$ denotes a pure state of a lattice of $N$ sites with local dimension $d$, so that the Hilbert-space dimension is $D\!=\!d^{N}$; for a bipartition $A\!\cup\!B$ we write $D\!=\!d_{A}d_{B}=d^{n_A+n_B}$, and without loss of generality $d_A \leq d_B$. The RDM is defined as $\rho_{A}\!=\!\Tr_{B}\dyad{\psi}$, and we denote by $\qty{\lambda_{i}}$ the entanglement spectrum, \idest{} the eigenvalues of $\rho_{A}$. 
Entropies are measured in bits unless stated otherwise.

\subsection{Global magic}
The free pure states of the resource theory of magic are the stabilizer states, \idest{} the Clifford orbit of the computational-basis product states~\cite{Gottesman1997StabilizerCodesQuantum, Gottesman1998TheoryFaulttolerantQuantum, Aaronson2004ImprovedSimulationStabilizer, Veitch2014ResourceTheoryStabilizer}, and the resource is measured relative to the generalized Pauli, or Heisenberg--Weyl, group
\begin{equation}
\mathcal{P}_{N}\!=\!\qty{\hat{P}\!=\!\bigotimes_{\site=1}^{N}\X^{p_{\site}}\Z^{q_{\site}}\Big|p_{1}\dots p_{N},q_{1}\dots q_{N}\!\in\!\mathbb{Z}_{d}},
  \label{eq_pauli_group}
\end{equation}
understood modulo phases, so that $\mathcal{P}_{N}$ consists of $D^{2}$ representatives. Here $\X$ and $\Z$ are the shift and clock operators, $\X\ket{\alpha}\!=\!\ket{\alpha\!\oplus\!1}$ and $\Z\ket{\alpha}\!=\!\omega^{\alpha}\ket{\alpha}$ with $\omega\!=\!e^{2\pi i/d}$ and $\oplus$ the sum modulo $d$; for $d\!=\!2$ they reduce to $\Sx$ and $\Sz$, and \cref{eq_pauli_group} to the usual Pauli group.

A pure state induces a probability distribution over $\mathcal{P}_{N}$, the Pauli-weight distribution
\begin{equation}
  \Xi_{\hat{P}}(\psi)=\frac{1}{D}\abs{\mel{\psi}{\hat{P}}{\psi}}^{2},
  \qquad \sum_{\hat{P}\in\mathcal{P}_{N}}\Xi_{\hat{P}}=\Tr\rho^{2}=1,
  \label{eq_pauli_weight}
\end{equation}
where $\rho$ is the global pure state.
The global magic can then be defined as the \textbf{Stabilizer-Rényi entropy} (SRE)~\cite{Leone2022StabilizerRényiEntropy, Leone2024StabilizerEntropiesAre}:
\begin{equation}
\begin{split}
  \mathcal{M}_{\text{SRE}}(\psi)&\!=-\log_{2}\!\!\qty[\!
    \frac{1}{D}\!\!\sum_{\hat{P}\in\mathcal{P}_{N}}\!\!\!
    \abs{\!\mel{\psi}{\hat{P}}{\psi}}^{4}]\\
  &=-\log_{2}\sum_{\hat{P}}\Xi_{\hat{P}}^{2}\!-\!\log_{2}\!D,
\end{split}
  \label{eq_SRE}
\end{equation}
which reads as the Rényi-$2$ entropy of $\Xi_{\hat{P}}$, offset by the stabilizer baseline $\log_{2}D$. For a stabilizer state, $\Xi_{\hat{P}}$ is uniform over the $D$ elements of the stabilizer group, exactly saturating the offset and yielding a vanishing SRE~\cite{Leone_2024,Haug2023QuantifyingNonstabilizernessMatrix}.
The $\mathcal{M}_{\text{SRE}}$ is efficiently measurable via Bell and Pauli sampling~\cite{Oliviero2022MeasuringMagicQuantum, Haug2023ScalableMeasuresMagic} and, for the special states considered here, admits an exact evaluation via the transfer-matrix methods developed in Appendices~\ref{app:SRE_Z2_scars} and~\ref{app:SRE_PXP_scar_exact}.

While the SRE provides an entropic measure of nonstabilizerness, geometric measures can also be defined. One such measure is the trace-distance magic:
\begin{align}
    \mathcal{M}_{\text{dist}}(\psi)=\underset{\sigma \in \text{STAB}_0^{(N)}}{\text{min} }\frac{1}{2}|| \rho_{\psi} - \sigma||_1,
\end{align}
with
\begin{align}
    &\text{STAB}_0^{(N)}:= \nonumber\\
    &\{ \sigma=\frac{1}{d^N} \sum_{P \in G} P \; \vert \; G \subset \mathcal{\tilde{P}}_N \; \text{a stabilizer group}\},
\end{align}
where $\tilde{\mathcal{P}}_N=\langle \tilde{\omega},X,Z \rangle$ is the Heisenberg-Weyl group including phases, with $\tilde{\omega}=\omega$ for odd $d$ and $\tilde{\omega}=e^{i \pi/2}$ for $d=2$. The entanglement spectrum of states in $\text{STAB}_0^{(N)}$ is flat with rank $d^m$ for integer $m$. $\text{STAB}_0^{(N)}$ also defines the set of states on which $\mathcal{M}_{\text{SRE}}$ vanishes. It is a strict subset of the full stabilizer polytope $\text{STAB}^{(N)}$. In the following, we only consider qubit systems, corresponding to local dimension $d=2$.

\subsection{Nonlocal magic}

\begin{figure}
    \centering
    \includegraphics[width=1.0\linewidth]{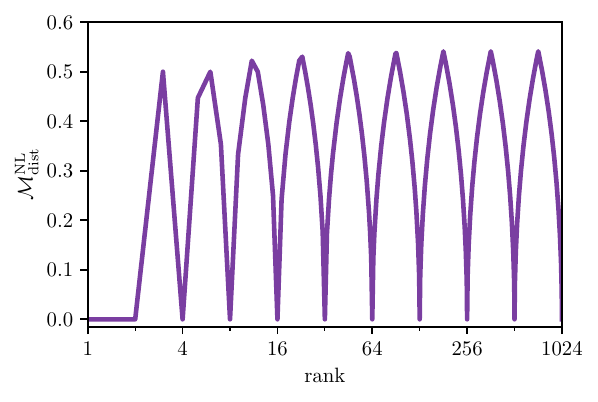}
    \caption{\textbf{Nonlocal magic from Schmidt-rank incompatibility.} Exact nonlocal trace-distance magic $M_{\mathrm{dist}}^{\mathrm{NL}}$
    for a flat entanglement spectrum of rank $p$. The measure vanishes when $p=2^m$ and is finite otherwise, showing that a perfectly flat
    spectrum can carry nonlocal magic solely because its Schmidt rank is incompatible with the dyadic Schmidt ranks allowed for pure stabilizer states.}
    \label{fig:maxiamlly_mixed_NL_dist}
\end{figure}

The global magic is basis dependent, and part of it might be undone by a change of local frame. 
What survives such changes is the nonlocal magic, obtained by minimizing a faithful magic measure $\mathcal{M}$ over local unitaries across the bipartition~\cite{Cao2025GravitationalBackreactionMagical},
\begin{equation}
    \mathcal{M}^{\text{NL}}(\psi_{AB})\!=\! \min_{U}\mathcal{M}(U\psi_{AB}U^{\dagger}),\; U\!=\!U_A\otimes U_B.
    \label{eq_NLM_def}
\end{equation}

In this paper, we consider two faithful measures of magic. The first is the nonlocal trace-distance magic, for which we derive an exact analytic expression determined solely by the Schmidt spectrum. In the following the entanglement spectrum is ordered in decreasing order $\lambda_1 \geq \lambda_2 \geq \dots \geq0$ and we assume $d_A \leq d_B$. For $i>r_A$, we set $\lambda_i:=0$. To evaluate the nonlocal magic exactly, we compare it explicitly to pure stabilizer states and show that no mixed stabilizer state can increase the distance. The formula reads:
\begin{align} \label{eq_mdist_def}
     \mathcal{M}^{\text{NL}}_{\text{dist}} (\psi_{AB})&:=\sqrt{1- \underset{K \in \{ 1,d,...,d^{n_A} \}}{\mathrm{max}}F_K},\nonumber\\
    F_K&=\left(\sum_{i=1}^{\text{min}(r_A,K)} \sqrt{\lambda_i(\rho_A)} \right)^2/K.
\end{align}

A detailed derivation is provided in Appendix~\ref{app:LW_UP_NL_magic}.

The second is the nonlocal Stabilizer-Rényi entropy $\mathcal{M}^{\text{NL}}_{\text{SRE}}$. It admits an upper bound obtained by rotating the Schmidt vectors onto eigenstates of a stabilizer group~\cite{Cao2025GravitationalBackreactionMagical,liu2026entanglement}. This is one specific choice of a local basis change and does not have to be the optimal one.

\begin{align}
\mathcal{M}_{\text{SRE}}^{\text{NL}}(\psi_{AB}) &\leq \mathcal{M}_{\text{SRE}}( \qty{ \lambda_i }), \nonumber\\
\mathcal{M}_{\text{SRE}}( \qty{ \lambda_i })&= \mathcal{M}_{\text{SRE}} \qty( \sum_{i=1}^{2^{\kappa}} \sqrt{\lambda_i} \ket{s_i}\ket{s_i} )\; \kappa:=\lceil \text{log}_2 r_A \rceil, \label{eq_NLSRE}
\end{align}
with $\lambda_i:=0$ for $r_A< i \leq 2^{\kappa}$, and $\{|s_i \rangle\}$ computational (Z-)basis states.

For both measures, the generally computationally demanding minimization in Eq.~\eqref{eq_NLM_def} can be circumvented using expressions that depend only on the entanglement spectrum.\footnote{One can also incorporate the constraint of the model in the measure to obtain a valid upper bound. In both $\mathcal{M}^{\text{NL}}_{\text{dist}}$ and $\mathcal{M}_{\text{SRE}}(\{ \lambda_i\})$, we can choose pure stabilizer reference states that obey the constraint. This corresponds to unitaries that conserve the constraint, $\left[ U_A \otimes U_B,C\right]=0$, where $C$ is the conserved quantity. The exact $\mathcal{M}^{\text{NL}}_{\text{dist}}$ then becomes a lower bound on the nonlocal magic under constraint-preserving unitaries.} 

\subsubsection{Constraint systems}

In the following, we look at constrained systems, where the structure of the physical Hilbert space restricts the admissible Schmidt ranks. The rank is bounded by the dimension of the physical Hilbert space of the subsystem, $\text{rank}\rho_A \leq \text{rank} \mathcal{H}_A^{\text{phys}}$. Generic (ergodic) eigenstates often satisfy the equality. For the $\mathbb{Z}_2$ gauge theory considered here, the physical subspace has dimension $2^{n_A}$ for a small enough subsystem and half Ising domain wall filling. For the PXP model the situation is different: the Rydberg blockade gives $\mathcal{H}_A^{\text{phys}}=F_{L_A+2}$, where $F_n$ denotes the $n$-th Fibonacci number ($F_1=F_2=1$, $F_n=F_{n-1}+F_{n-2}$), and since the only Fibonacci numbers that are powers of $2$ are $2$ and $8$, the rank can be a power of two for $L_A=1$ and $L_A=4$. For U$(1)$ and SU$(2)$ gauge theories the physical Hilbert-space dimension is likewise not generally a power of $2$, except in some special cases such as group truncations that permit a mapping of the physical Hilbert space to a spin chain Hilbert space.

The consequences are illustrated for a flat spectrum $\lambda_i=\frac{1}{p}$ for $i=1,\ldots,p$, where $p$ is the rank. The $\mathcal{M}_{\text{dist}}^{\text{NL}}$ takes the form

\begin{align}
    \mathcal{M}_{\text{dist}}^{\text{NL}}&=\sqrt{1-F(p)},\nonumber\\
    F(p)&=\text{max}(2^m/p,p/2^{m+1})
\end{align}
with $m=\lfloor \text{log}_2 p\rfloor $. In Fig.~\ref{fig:maxiamlly_mixed_NL_dist}, we show $\mathcal{M}_{\text{dist}}^{\text{NL}}$. The two arguments of the maximum are the two ways of matching a rank-$p$ system to a stabilizer one: truncating to the largest power of $2$ below $p$, or embedding into dimension $2^{m+1}$ and allowing $2^{m+1}-p$ zero eigenvalues. $\mathcal{M}_{\text{dist}}^{\text{NL}}$ vanishes exactly at $p=2^m$ and is strictly positive otherwise, reaching at most $\sqrt{1-1/\sqrt{2}} \approx 0.54$ near $p = \lfloor 2^m\sqrt{2} \rceil$. A perfectly flat spectrum can therefore produce nonlocal magic purely because its rank is incompatible with the stabilizer structure. The measure therefore responds to two distinct sources of nonlocal magic: spectral nonflatness and stabilizer-incompatible Schmidt rank.

This propagates to the global magic on $\text{STAB}_0$, since for every bipartition $\mathcal{M}^{\text{NL}}(\rho_A) \leq \mathcal{M}(\psi)$ and hence
\begin{align}
   \underset{A|B}{\text{max}} \mathcal{M}^{\text{NL}}(\psi) \leq \mathcal{M}(\psi). 
\end{align}

In Secs.~\ref{sec:Dynamics} and~\ref{sec:statics}, we first consider cases in which the maximal possible rank of the reduced state is a power of $2$. In Sec.~\ref{sec:not_power_2}, we examine how nonlocal magic changes when the accessible Schmidt rank is not a power of $2$.

\section{Models}
Our characterization of quantum resources is performed on two different models known to exhibit scars: the PXP model and a $\mathbb{Z}_2$ gauge theory.

\textit{The PXP model.} The PXP Hamiltonian with periodic boundary conditions is defined as:
\begin{equation}
    \ham^{\text{PB}}_{\mathrm{PXP}}=\sum_{\site=1}^{L}\hat{P}_{\site-1} \Sx_{\site} \hat{P}_{\site+1},
    \label{eq_PXP}
\end{equation}
where $\hat{P}_{\site}=(\mathbb{1}-\Sz_{\site})/2$ and the sites $j$ and $j+L$ are identified with each other. For open boundary conditions the Hamiltonian reads:
\begin{align}
    \ham^{\text{OB}}_{\mathrm{PXP}}=X_1P_2+\sum_{\site=2}^{L-1}\hat{P}_{\site-1} \Sx_{\site} \hat{P}_{\site+1}+P_{L-1}X_L.
\end{align}
The Hamiltonian decomposes into blocks labeled by the conserved quantity $\hat{N}_{\uparrow\uparrow}\!=\!\sum_{\langle\site,\site+1 \rangle}(1\!+\!\Sz_{\site})(1\!+\!\Sz_{\site\!+\!1})/4$, which counts the pairs of neighboring excitations. It admits a one-to-one mapping with a U$(1)$ Schwinger model \cite{Surace2020LatticeGaugeTheories}.
We focus on the blockade-satisfying sector $N_{\uparrow\uparrow}\!=\!0$. 
The overall system exhibits quantum chaotic features, although deviations from perfect ergodicity occur at smaller system sizes, see Appendix~\ref{app:ergodicity}. 
This model hosts well-known scarred initial states, such as the Néel state, whose dynamics evade conventional thermalization despite its mean energy lying at the center of the spectrum \cite{Turner2018WeakErgodicityBreaking}.
The Néel state, also referred to as the period-2 charge-density-wave (CDW1) state, consists of alternating spin-up and spin-down spins. Translating this state by one spin yields the symmetry-related CDW2 state, which is likewise a scarred initial state. The Néel state has large overlap with special eigenstates. The entanglement entropy of these special scar eigenstates grows only logarithmically with system size and can be described by the forward scattering approximation~\cite{Turner_2018}. The PXP model also hosts exact scar states that persist in the thermodynamic limit for both OBC and PBC. These states lie in the middle of the spectrum and exhibit area-law entanglement scaling~\cite{Lin2019ExactQuantumManyBody, Surace_2021}. In the following, we restrict our analysis to the $k=0$ and $k=\pi$ momentum sectors under PBC, which are the only sectors with nonzero overlap with the Néel state.

\textit{The $\mathbb{Z}_2$ model.}
Another Abelian model, introduced in \cite{Iadecola2020QuantumManybodyScar}, is given by the following Hamiltonian:
\begin{equation}
    \ham\!=\!\sum_{\site=2}^{N-1}\qty[\lambda( \Sx_{\site} - \Sz_{\site-1} \Sx_{\site} \Sz_{\site+1})+ \Delta \Sz_{\site} + J\Sz_{\site} \Sz_{\site+1} ].
    \label{eq_Z2_hamiltonian}
\end{equation}
It has a global U$(1)$ symmetry whose associated conserved charge is the number of Ising domain walls $\hat{n}_{\mathrm{DW}}=\sum_{\langle \site,\site+1 \rangle}[1-\Sz_{\site} \Sz_{\site+1}]/2$, with $[\hat{n}_{\mathrm{DW}},\hat{H}]=0$. We consider OBC with the two edge spins frozen to $(\sigma^z_1,\sigma_L^z)=(-1,-1)$.
It can be mapped to the model studied in~\cite{Halimeh2023RobustQuantumManybody}, which carries a local $\mathbb{Z}_2$ gauge symmetry. 
For the parameters we consider, $\lambda=1$, $J=1$, and $\Delta=0.1$, the model is ergodic, and most of its eigenstates satisfy the ETH, see Appendix~\ref{app:ergodicity}. Nevertheless, it hosts a finite number of exactly solvable QMBS states $\ket{S_n}$ that display nonthermal behavior, where $n$ denotes the number of magnons. The scar states read:
\begin{align}
    |\mathcal{S}_n\rangle = \frac{1}{\mathcal{N}} 
    \sum_{\substack{\text{configs. with }\\ n \text{ excitations,} \\ 
    \text{no two adjacent}}} 
    (-1)^{\sum_{p} i_p} \;|0 \cdots \underset{i_1}{1}  \cdots \underset{i_n}{1} \cdots 0\rangle,
    \label{eq:Z2_scar_def}
\end{align}
where $i_p$ is the position of the $p$-th excitation,
and $\mathcal{N} = \sqrt{\binom{N-n-1}{n}}$. These scar states also admit an MPS representation~\cite{Iadecola2020QuantumManybodyScar}.

\begin{figure*}
    \includegraphics[width=1.0\textwidth]{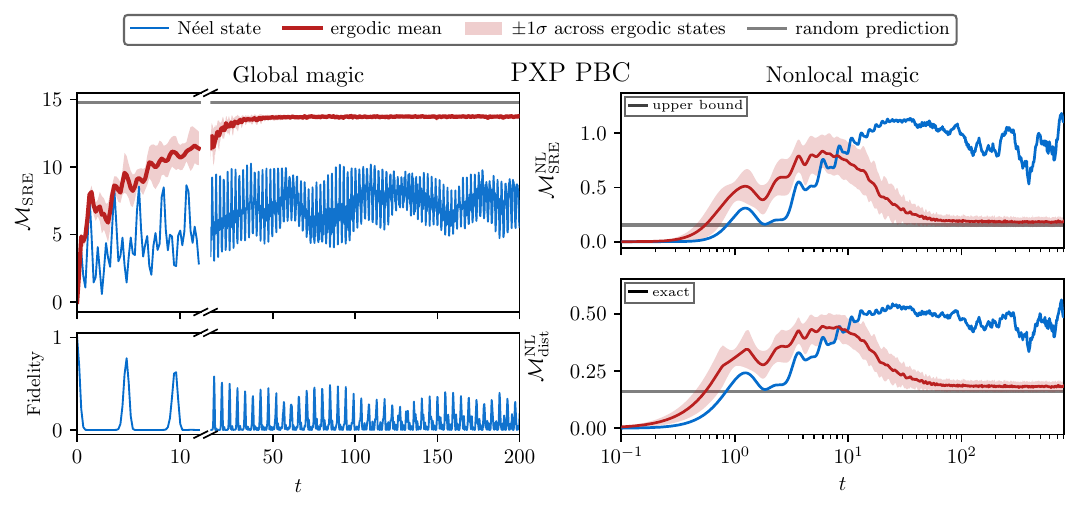}
    \caption{\textbf{Quantum-resource dynamics in the PXP model with PBC.} Time evolution starting from the Néel state, compared with the mean over ergodic states at the same energy. Left: global SRE, the bottom figure shows the corresponding fidelity revivals. Right: nonlocal magic, quantified by the upper bound on $\mathcal{M}^{\text{NL}}_{\text{SRE}}$ (top) and the exact $\mathcal{M}^{\text{NL}}_{\text{dist}}$ (bottom). The total system size is $N=18$ and the subsystem size $n_A$ is $4$ qubits. Shaded regions indicate the $\pm 1\sigma$ fluctuations across ergodic states. The gray reference denotes the corresponding random-state prediction. While the global magic exhibits pronounced oscillations correlated with the scar revivals, the nonlocal magic remains anomalously large along the scarred trajectory compared with the ergodic behavior.}
    \label{fig:PXP_dynamics}
\end{figure*} 

\section{Results}
All results in this section follow the same protocol.
We first compute, numerically, the magic measure for an ensemble of $50$ random states, respecting the constraints of each model in the overall $2^N$ Hilbert space spanned by the spin chain. 
For measures based on the real eigenvectors of the Hamiltonian, we draw real random numbers, whereas for measures based on the time-evolved state we draw complex random values, owing to the complexity introduced by $e^{-iHt}$. 
We then compute the same magic measure for the QMB scars and an ensemble of ergodic states with an energy close to the QMB scars. We set the energy window for the ergodic states to be: $|E-E_{\text{scar}}|\leq 1/8$. Finally, we compare the ergodic result with the random prediction to identify deviations in the QMBS. For the dynamics, we evolve $20$ initial product states with the same energy as the Néel state, which are not scarred initial states.

We divide the discussion into the dynamical and static characterization of QMB scars, presented in Secs.~\ref{sec:Dynamics} and~\ref{sec:statics}, respectively. 
In both cases, we examine the global SRE and the nonlocal magic across the different models.

\subsection{Scarring dynamics of quantum resources}\label{sec:Dynamics}

\textit{Global magic.} 
To characterize the dynamics, we compare the time evolution of the SRE with that of the fidelity $\mathcal{F}(t)\!=\!\abs{\braket{\psi(0)|\psi(t)}}^2$ for a product initial state. As initial scarred state, we choose the Néel state~\cite{Surace2020LatticeGaugeTheories}. The results for a system size of $N=18$ are shown in the left column of Fig.~\ref{fig:PXP_dynamics}.
\begin{figure*}
    \centering
    \includegraphics[width=1.0\linewidth]{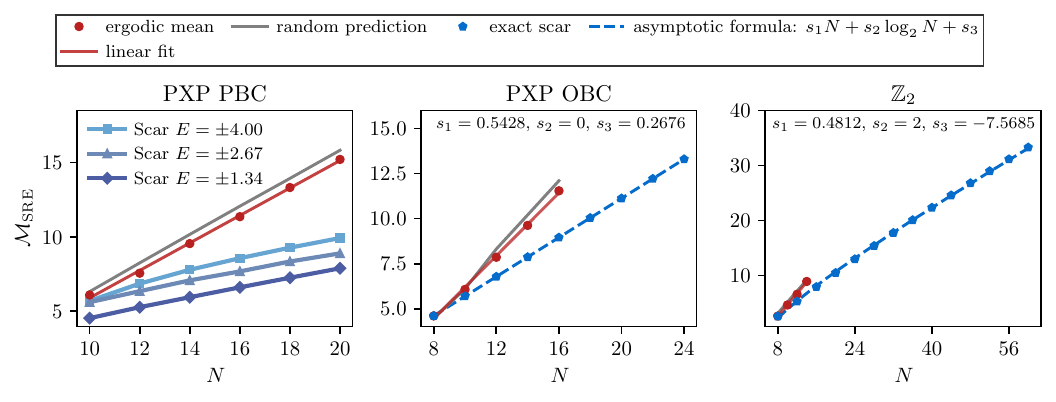}
    \caption{\textbf{Global magic of scarred and ergodic eigenstates.} System-size scaling of the SRE of scarred eigenstates and nearby ergodic eigenstates in the PXP model with periodic (left) and open (middle) boundary conditions, and in the $\mathbb{Z}_2$ model (right). For the PXP model with PBC, several scar states at different energies are shown; for OBC, the exact scar states $|\Gamma_{12}\rangle$ and $|\Gamma_{21}\rangle$ are compared with the analytical large$-N$ result. For the $\mathbb{Z}_2$ model, the exact scar $|S_{N/4}\rangle$ SRE is compared with the asymptotic form $s_1N+s_2\log_2N+s_3$. Ergodic eigenstates exhibit the expected extensive scaling and approach the random prediction. In all cases, scar states also display an extensive global magic, but with a smaller leading slope than their ergodic counterparts.}
    \label{fig:SRE_static}
\end{figure*}
We observe clear fidelity revivals, which correspond to periodic recurrences of the initial state. The same behavior is mirrored in the SRE, which starts at $0$, since the initial state is a product state of single-qubit stabilizer states and thus carries no magic. 
Then, the state spreads into superpositions and acquires magic. 
The SRE reaches a minimum when the CDW1 state transitions into the CDW2 state, and vice versa, before returning close to its initial value/state. 
This SRE behavior can already be partially deduced from the fidelity, which, as proven in Appendix~\ref{app:Fidelity_SRE_relation}, provides the following upper bound on the SRE:
\begin{equation}
    \SRE(t)\leq -4\log_2\left[\mathcal{F}(t)\right].
    \label{eq_M2_fidelity}
\end{equation}
Although this bound is generally very loose, it becomes informative for states close to product states.
It implies that, for fidelities close to $1$, the SRE is close to $0$.
Over long times, the fidelity exhibits periodic peaks for the PXP model with values between $\sim\!0.2$ and $\sim\!0.5$. At these points, the SRE is therefore bounded from above by magic values between $\sim\!4.0$ and $\sim\!9.3$. This is a clear deviation from the ergodic behavior, for which the SRE closely approaches the Haar-random value $\SRE^{\rm Haar}\! \approx \!14.8$.
Thus, the global SRE is a faithful witness of scarring dynamics for initial scarred product states: it oscillates in phase with the return fidelity and, crucially, never approaches the value expected for a typical ergodic state, whose magic, conversely, saturates. 
These oscillations do not constitute an independent signature of scarring: \cref{eq_M2_fidelity} shows that a fidelity revival necessarily constrains the SRE from above, so the observed dips are, to a large extent, a consequence of the revivals themselves.
What the SRE adds beyond the fidelity is the quantitative statement that the magic generated along the scarred trajectory remains below the typical value at all times. 

\textit{Nonlocal magic.} The global SRE is basis-dependent and can change if the model is expressed in a different basis. 
To estimate how much magic cannot be removed by local basis changes, we turn to the nonlocal magic. 
For a chain of size $N$, we focus on a subsystem of $n_A=4$ qubits, deliberately choosing a subsystem whose maximal rank is a power of $2$.

The upper bound on the nonlocal SRE and the exact nonlocal trace-distance magic for the PXP model are shown in the right column of Fig.~\ref{fig:PXP_dynamics}.
For the ergodic ensemble, all values start at zero, reach a peak, and then return to a small value, which is predicted by a random state (gray lines). A similar behavior was previously observed in the SU(2) model using a lower bound on $\mathcal{M}_{\text{dist}}^{\text{NL}}$~\cite{Ebner2026MagicBarrierThermalization}. Here, the exact $\mathcal{M}_{\text{dist}}^{\text{NL}}$ confirms that this barrier-like structure persists for the full nonlocal trace-distance magic.
In contrast, for the Néel state, the nonlocal magic likewise rises rapidly to a value similar to that of the ergodic state, but, unlike the latter, it does not drop down to a small value: it instead remains at a relatively high value. 
The entanglement spectrum of the time-evolved scarred state therefore remains strongly nonflat and distinct from the random-state prediction. We also note that the rank starts at $1$ and increases during the time evolution, passing through different intermediate values before eventually reaching the full possible rank at long times (here a power of $2$).
For the $\mathcal{M}_{\text{SRE}}^{\text{NL}}$, we can also compare the global magic on the left to the nonlocal upper bound on the right. The comparison indicates that most of the global magic is removable by local basis transformations.

These observations can be understood as follows: the CDW1 (or CDW2) initial state lies exactly at the mean energy of the model, corresponding via $E=\text{Tr}\left[He^{- \beta H}\right]/Z$ to infinite temperature $T\!=\!\infty$. 
If subsystem ETH holds, the RDM of a small subsystem $A$ approaches its thermal form once the bath $B$ is sufficiently large; at $T\!=\!\infty$, this is the maximally mixed state, which has, by definition, $\mathcal{M}_{\text{SRE}}^{\text{NL}}(\psi_{AB})=\mathcal{M}^{\text{NL}}_{\text{dist}}(\psi_{AB})=\!0$; finite-size and constraint-induced corrections can produce deviations from this ideal limit. For our $4$-qubit subsystem, this is a very rough approximation, especially with the PXP constraint at the boundaries. We attribute the residual nonlocal magic primarily to finite-size and constraint-induced boundary effects, consistent with the analogous deviations from zero observed for the random-state prediction.
The nonthermal nature of the scarred dynamical trajectory is precisely its failure to relax to this random/ETH prediction: the evolved state remains locally non-generic and retains a structured entanglement spectrum, while ergodic states become locally approximately thermal, which, here at $\beta \approx 0$, is structureless, and are captured by the universal random-state description.

\subsection{Quantum resources of scarred eigenstates}\label{sec:statics}
Let us now consider the scaling of global and nonlocal magic measurements versus system size $N$ for scar Hamiltonian eigenstates in comparison with standard ergodic ones. 
For this analysis, we focus on the PXP model in \cref{eq_PXP} and on the $\mathbb{Z}_{2}$ gauge theory in \cref{eq_Z2_hamiltonian}, for which analytical expressions of scarred eigenstates are known (see Appendix~\ref{app:SRE_PXP_scar_exact} and \cite{Smith2025NonstabilizernessKineticallyConstrained, Lin2019ExactQuantumManyBody} for the PXP model, \cite{Iadecola2020QuantumManybodyScar} and Appendix~\ref{app:SRE_Z2_scars} for the $\mathbb{Z}_{2}$ model) and therefore allow for an extensive study in $N$. 
The comparison is performed with corresponding scaling of an ergodic state close to the scar energy.

\begin{figure*}
    \centering
    \includegraphics[width=1.0\linewidth]{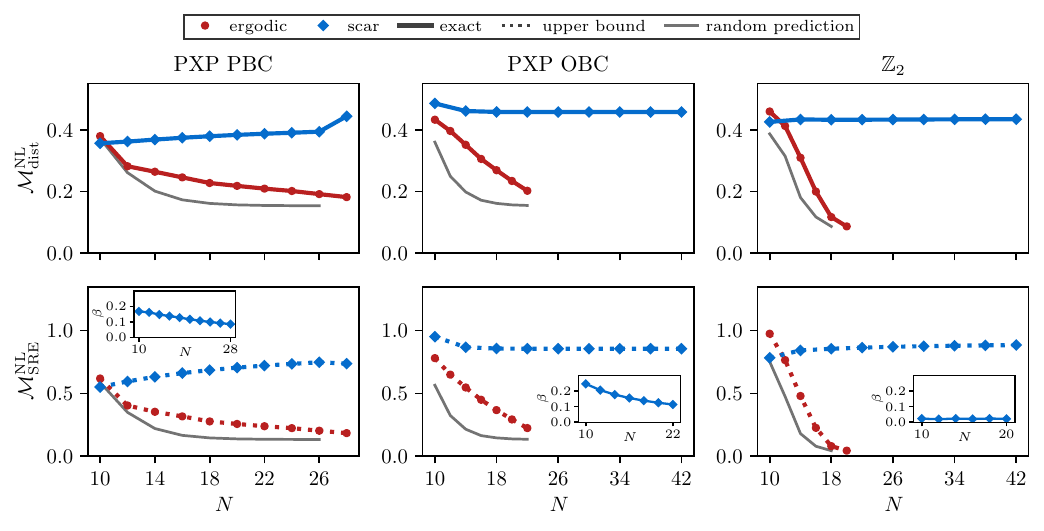}
    \caption{ \textbf{Nonlocal magic of scarred and ergodic eigenstates.} System-size scaling of the nonlocal magic for the same scar and nearby ergodic eigenstates in the PXP model with periodic (left) and open (middle) boundary conditions, and in the $\mathbb{Z}_2$ model (right). The states considered are the same as in Fig.~\ref{fig:SRE_static}. The subsystem $n_A$ is fixed to $4$ qubits. Top: exact nonlocal trace distance magic $M_{\mathrm{dist}}^{\mathrm{NL}}$. Bottom: upper bound on the nonlocal stabilizer Rényi entropy $M_{\mathrm{SRE}}^{\mathrm{NL}}$. The insets in the bottom panels show the inverse temperature $\beta=1/T$ of the scar states as a function of the system size. While the nonlocal magic of ergodic eigenstates decreases toward the random-state expectation as the environment grows, scar eigenstates retain a substantially larger nonlocal resource, reflecting the persistent non-flat structure of their entanglement spectrum.}
    \label{fig:Nl_magic_static}
\end{figure*}

\textit{Global magic.} 
For the PXP model with open boundary conditions, there exist four scar states $\ket{\Gamma_{\alpha\beta}}$, with $\alpha,\beta\in\qty{1,2}$, which admit an exact bond-dimension-$2$ MPS representation (see Appendix~\ref{app:SRE_PXP_scar_exact}). 
We focus on $\ket{\Gamma_{12}}$, $\ket{\Gamma_{21}}$
, which have definite symmetry quantum numbers and are located at energies $E=E_{\mathrm{mean}}\pm\sqrt{2}$, where $E_{\mathrm{mean}}$ is always zero in the PXP model. As the scar energy density corresponds to infinite temperature in the thermodynamic limit and follows area-law entanglement, scars severely violate the strong ETH.
Away from the boundaries, these states are identical in every two-site unit cell. As the Pauli strings factorize over unit cells as well, the SRE can be evaluated exactly by transfer-matrix techniques. 
In the large-$N$ limit, we find a volume law scaling:
\begin{equation}
    \SRE(\Gamma)\approx
     0.542817\,N+0.2676.
\end{equation}
The derivation can be found in Appendix~\ref{app:SRE_PXP_scar_exact}. The scar result together with the mean ergodic states result is shown in the middle of Fig.~\ref{fig:SRE_static}. From the numerical results up to $N=16$ we observe that the typical ergodic state grows linearly as expected and closely approaches the random-state prediction. The SRE of both the scar and the ergodic state increases linearly, but the latter with a larger slope than the former.

We repeat the same analysis for the scar eigenstates in the PBC PXP model, which cause the dynamical Néel state oscillations in Fig.~\ref{fig:PXP_dynamics}. We show the three scar states with the largest  Néel state overlap in  Fig.~\ref{fig:SRE_static}. Because these scar eigenstates do not permit an exact analytic description, we are limited by system size. However, for the system sizes we can access, we see the same behavior as for the exact OBC scar state. For the accessible system sizes, the mean ergodic SRE increases linearly, while the scar states appear to exhibit linear scaling with a smaller slope, although logarithmic scaling cannot be ruled out conclusively.

We extend the analysis to the $\mathbb{Z}_2$ gauge theory in \cref{eq_Z2_hamiltonian}. We focus in particular on the scar state $\ket{S_{n}}$ which contains $n=N/4$ magnons; since each magnon is an isolated flipped spin, it carries two Ising domain walls, and the state lies in the symmetry sector with $n_{\mathrm{DW}}=N/2$. It is an equal-weight superposition of $\binom{3N/4-1}{N/4}$ basis states of this sector. The scar state is written down explicitly in Eq.~\eqref{eq:Z2_scar_def}.
This specific scar state has logarithmic entanglement entropy scaling shown in Ref.~\cite{Iadecola2020QuantumManybodyScar}. For this scar state we find the following analytic large-$N$ formula for the SRE:
\begin{align}
    \mathcal{M}_{\text{SRE}}(|S_{N/4}\rangle)\approx 0.4812 N+2 \mathrm{log}_2(N)+C. 
\end{align}
The derivation can be found in Appendix~\ref{app:SRE_Z2_scars}. If we do not apply the large-$N$ approximation, we can calculate the SRE for the scar state exactly up to $N=56$. With these results we fix the constant $C=-7.5685$. The comparison between the exact scar SRE value (blue pentagons), the asymptotic formula (dashed blue line) and the ergodic state is shown in the right panel of Fig.~\ref{fig:SRE_static}. We again observe the same behavior, with both showing a leading linear scaling.

Combining what we found for the PXP and the $\mathbb{Z}_2$ model, we conclude that scarred and ergodic eigenstates need not exhibit parametrically different scaling laws of global magic: both can display volume-law nonstabilizerness, although with different leading coefficients. This stands in marked contrast to their entanglement entropy, for which scarred eigenstates exhibit parametrically reduced scaling. Nevertheless, the fact that global nonstabilizerness and entanglement are smaller for scar states than for ergodic states is a shared feature between SRE and entanglement.

\textit{Nonlocal magic.}
We now turn to nonlocal magic, repeating the analysis on the same scarred eigenstates of the three models. We choose the subsystem to be $4$ qubits centered in the middle of the chain such that the numbers of spins to its left and right are both odd.\footnote{This choice is relevant for the PXP OBC model. When the subsystem is instead surrounded by an even number of spins on both sides, the nonlocal magic is anomalously small, falling below the random-state prediction.} This means that the rank of all ergodic states considered is a power of $2$. We again consider the exact $\mathcal{M}^{\text{NL}}_{\text{dist}}$ and the upper bound on $\mathcal{M}_{\text{SRE}}^{\text{NL}}$. The results are shown in Fig.~\ref{fig:Nl_magic_static}. For the PBC scar states we show for visual clarity only the one scar state at $E=\pm 1.33$ out of the three considered in Fig.~\ref{fig:SRE_static}. The other two scar states behave in a similar fashion. In all three models, the mean ergodic nonlocal magic decreases to a small number. The $\mathbb{Z}_2$ curve is the closest to the random prediction. For the PXP model, the ergodic mean approaches the random prediction for larger system sizes. The discrepancy between the random prediction and the ergodic state for smaller $N$ in the PXP model can be explained via the slow approach of $\beta$ towards $0$. For the $\mathbb{Z}_2$ model the scar state lies almost exactly at $\beta \approx0$ from the beginning.  Furthermore, the PXP model random ergodic result approaches a small but finite number. This fact we attribute as before to the PXP constraint across the boundary. All three scar examples show that the nonlocal magic of the scars stays large and finite. For the PXP OBC and the $\mathbb{Z}_2$ theory this statement is analytically well established in the thermodynamic limit. For the PXP scars (OBC and PBC), the rank of the scar is a power of $2$, such that the measure genuinely reflects the antiflat structure of the spectrum. For the $\mathbb{Z}_2$ model, the scar state always has rank $5$ or $7$ across the shown data points. Thus, the high nonlocal magic can be attributed not only to the nonflat spectrum, but also to the incompatibility of the rank with that of a stabilizer state.
Overall, we see a clear difference between the ergodic states, which converge to the random prediction and the scar states. It is important to keep in mind that the energies considered here correspond to infinite temperature $\beta=0$, in the thermodynamic limit. For a nonzero $\beta$, the entanglement spectrum develops nonflatness, giving rise to nonlocal magic. Consequently, the large $n_B$ spectrum of the subsystem $n_A$ may no longer be adequately described quantitatively by a structureless random-matrix prediction alone. 

Taken together, these results reveal a qualitative separation between global and nonlocal magic. Global magic grows extensively for both scarred and ergodic eigenstates and therefore does not parametrically distinguish the two. By contrast, for the fixed subsystems considered here, nonlocal magic is suppressed toward the random-state expectation as the environment grows for ergodic eigenstates, while remaining anomalously large for scarred eigenstates.

\section{Nonlocal magic for non-dyadic rank}\label{sec:not_power_2}
In this section, we examine the consequences of non-dyadic rank for the nonlocal magic of a subsystem. In addition to spectral nonflatness, a second source of nonlocal magic then arises from the incompatibility of the Schmidt rank with the dyadic ranks allowed for stabilizer states. For the PXP dynamics, we consider a subsystem of $6$ qubits, for which the RDM has rank $21$. The example is intended to illustrate the effect rather than to provide an exhaustive characterization. For both the $\mathcal{M}_{\text{dist}}^{\text{NL}}$ and the upper bound on $\mathcal{M}_{\text{SRE}}^{\text{NL}}$, the long-time value of the nonlocal magic is shifted upward relative to the dyadic-rank case. Qualitatively, however, the overall behavior remains largely unchanged, although the ergodic barrier form is less pronounced. The ergodic state still exhibits a peak before relaxing towards the random-state prediction, which lies at a substantially higher value. In contrast, the scar state does not relax towards this value; instead, it remains at a value of nonlocal magic that is consistently higher than that of the ergodic state.
\begin{figure}
    \centering
    \includegraphics[width=1.0\linewidth]{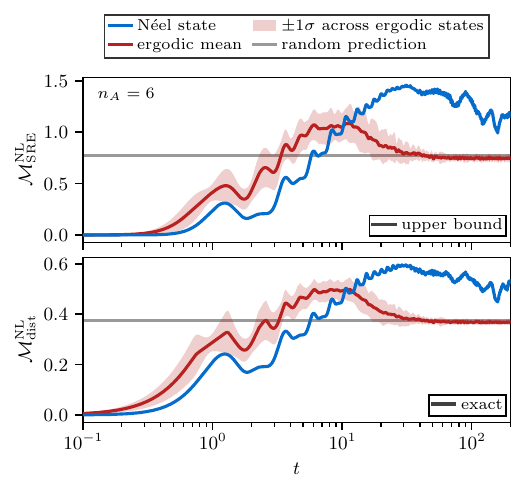}
    \caption{\textbf{Nonlocal magic dynamics for non-dyadic rank.} Time evolution starting from the Néel state, compared with the mean over ergodic states at the same energy. Top: nonlocal magic quantified by the upper bound on $\mathcal{M}_{\text{SRE}}^{\text{NL}}$. Bottom: exact nonlocal trace-distance magic $\mathcal{M}_{\text{dist}}^{\text{NL}}$. The subsystem size $n_A$ is $6$ qubits. The rank of the reduced density matrix is $21$, and thus not a power of $2$. }
    \label{fig:placeholder}
\end{figure}

\section{Conclusion}
In this work, we have investigated quantum many-body scars through the lens of nonstabilizerness, comparing global and nonlocal notions of magic in constrained many-body systems. We considered scarred dynamics and eigenstates in Abelian lattice gauge theories,
and contrasted their resource content with that of typical ergodic states. Our results show that the nonstabilizerness provides information complementary to standard diagnostics of weak ergodicity breaking, but that the distinction between global and genuinely nonlocal resources is important.

At the global level, the stabilizer Rényi entropy provides a clear dynamical signature of scarring. Starting from scarred product states, its oscillations track the fidelity revivals and remain well below the values characteristic of ergodic dynamics. For scarred eigenstates, however, the situation is qualitatively different from entanglement.

Despite their anomalously low entanglement, the exact scars of both the PXP and $\mathbb{Z}_2$ models exhibit an extensive global magic. In particular, we obtained analytical large-system expressions for the exact OBC PXP scars and for the Iadecola–Schecter $\mathbb{Z}_2$ scars. Thus, unlike entanglement entropy, the scaling of global nonstabilizerness does not parametrically distinguish scarred from ergodic eigenstates: both can obey a volume law, although with different coefficients.

A sharper distinction emerges after removing the magic that can be eliminated by local basis transformations. For bipartite pure states, the resulting nonlocal magic is completely determined by the Schmidt spectrum and therefore directly probes deviations from the locally thermal behavior expected under ETH. 
For fixed subsystems, ergodic states close to $\beta \approx 0$ approach the random-state prediction and their nonlocal magic is strongly suppressed as the environment grows. Scarred states instead retain a structured, nonflat entanglement spectrum and consequently an anomalously large nonlocal magic relative to the corresponding ergodic states. In this sense, the global and nonlocal measures reveal two complementary aspects of scarring: scar states can remain extensively magical as many-body states while simultaneously carrying an anomalous, irreducible resource encoded in their entanglement structure.
Constrained Hilbert spaces provide an additional ingredient absent in unconstrained spin systems. Because stabilizer states admit only dyadic Schmidt ranks, even a perfectly flat entanglement spectrum carries nonzero nonlocal magic whenever its rank is incompatible with this stabilizer structure. This establishes a direct connection between kinematic constraints and irreducible quantum resources, which is particularly relevant for lattice gauge theories and Rydberg-blockaded systems.
A complementary analytical result of this work is the exact characterization of the nonlocal trace-distance magic for pure multiqubit states. We show that the optimization over local unitary transformations and over stabilizer-code states can be reduced to a finite optimization determined solely by the Schmidt spectrum. In particular, the result is controlled by the overlap with flat, stabilizer-compatible Schmidt spectra of dyadic rank. This closed-form expression provides a direct link between nonlocal magic and entanglement-spectrum structure, and makes the measure accessible without performing the otherwise costly optimization over local bases.

\paragraph*{Note added.—}
While this manuscript was being finalized, two independent works
appeared~\cite{sierant2026exact,viscardi2026nonlocal} deriving exact
closed-form expressions for nonlocal magic based on the stabilizer
fidelity. In both cases, the optimization over local unitaries is shown
to depend only on the Schmidt spectrum and to reduce to an optimization
over the allowed dyadic stabilizer ranks. These results are closely
related to Eq.~\eqref{eq_mdist_def}: the maximal fidelity entering our expression is the same spectral quantity, while here we formulate the corresponding nonlocal trace-distance magic and apply it to quantum many-body scars and constrained Hilbert spaces. The physical applications considered here, including the role of non-dyadic Schmidt ranks, are complementary to those works.

\section{Acknowledgments}
L.E., G.C., K.X.~and J.C.H.~acknowledge funding by the Max Planck Society, the Deutsche Forschungsgemeinschaft (DFG, German Research Foundation) under Germany's Excellence Strategy – EXC-2111 – 390814868, and the European Research Council (ERC) under the European Union's Horizon Europe research and innovation program (Grant Agreement No.~101165667)—ERC Starting Grant QuSiGauge. This work is part of the Quantum Computing for High-Energy Physics (QC4HEP) working group. J.C.H.~acknowledges the hospitality of the Aspen Center for Physics while this paper was being completed.
E.T.~is funded by the Swiss National Science Foundation (SNSF) under Grant No. TMPFP2\_234754. E.T.~acknowledges the CINECA (Consorzio Interuniversitario per il Calcolo Automatico) award, under the ISCRA initiative and Leonardo early access program, for the availability of high-performance computing resources
and support.

\newpage

\appendix

\section*{Supplemental Material}
\section{Proof of the exact nonlocal trace distance magic}\label{app:LW_UP_NL_magic}

Throughout the proof, $|\psi_{AB}\rangle \in \mathcal{H}_A \otimes \mathcal{H}_B$ is a pure state of $n=n_A+n_B$ qubits. The RDM is $\rho_A=\text{Tr}_B(\rho)$ with $\rho=|\psi_{AB}\rangle \langle \psi_{AB} |$. We define $d_A=2^{n_A}$ and the rank (number of non-zero Schmidt values) as $r_A=\text{rank}\rho_A$. We assume $d_A \leq d_B$. We write the Schmidt decomposition as:
\begin{align}
    | \psi _{AB} \rangle = \sum_{i=1}^{r_A} \sqrt{\lambda_i} | \phi _i \rangle_A \otimes | \chi_i \rangle _B, \; \lambda_1 \geq ... \geq \lambda_{r_A} >0.
\end{align}
For $i>r_A$, we set $\lambda_i:=0$. 
The trace distance magic is defined via:
\begin{align}
    \mathcal{M}_{dist}(\rho)&= \underset{\sigma \in \text{STAB}_0^{(N)}}{\mathrm{min}} \frac{1}{2} || \rho -\sigma ||_1,
\end{align}
where $\text{STAB}_0^{(N)}$ denotes the set of normalized projectors onto stabilizer codes. Explicitly, every $\sigma \in \text{STAB}_0^{(N)}$ can be written as:
\begin{align}
    \sigma=\frac{P}{R},
\end{align}
where $P$ is the projector onto a stabilizer code of dimension:
\begin{align}
    R=2^r, \; r=0, \dots N.
\end{align}
The case $R=1$ corresponds to a pure stabilizer state. The nonlocal trace distance magic across the partition $A|B$ reads:
\begin{align}\label{eq_ap_nonlocal_trm}
\mathcal{M}_{\text{dist}}^{NL}=\underset{U_A,U_B}{\text{min}} \mathcal{M}_{\text{dist}} \left[ (U_A \otimes U_B) |\psi \rangle \langle \psi | (U_A^{\dagger} \otimes U_B^{\dagger})\right],
\end{align}
where $U_A$ and $U_B$ are arbitrary unitaries acting on the two subsystems. Our aim is to show that this quantity is determined exactly by the Schmidt spectrum.

The first point that needs to be established is that, although the minimization in Eq.~\eqref{eq_ap_nonlocal_trm} contains mixed stabilizer-code states, for a pure input state the optimum can always be attained by a pure stabilizer state. 

Consider an arbitrary mixed state $\sigma\in\mathrm{STAB}_0^{(N)}$. Such a state can be written as a normalized projector onto a stabilizer code,
\begin{equation}
    \sigma_{\mathcal C}=\frac{\Pi_{\mathcal C}}{R},
\end{equation}
where $\Pi_{\mathcal C}$ projects onto a stabilizer code
$\mathcal C$ of dimension $R=2^m$. For $R=1$ the state is already a pure stabilizer state, so we only need to consider $R>1$.
For a pure target state $|\psi\rangle$, define
\begin{equation}
q=\langle\psi|\Pi_{\mathcal C}|\psi\rangle, \qquad 0\leq q\leq 1.
\end{equation}
For $q>0$, we can decompose the state as
\begin{equation}
|\psi\rangle=\sqrt{q}\,|u\rangle+\sqrt{1-q}\,|v\rangle,
\end{equation}
where $|u\rangle\in\mathcal C$ and
$|v\rangle\perp\mathcal C$ are normalized. In the two-dimensional
subspace spanned by $|u\rangle$ and $|v\rangle$, the operator
$|\psi\rangle\langle\psi|-\Pi_{\mathcal C}/R$ is represented by
\begin{equation}
    \begin{pmatrix}
        q-\frac{1}{R} & \sqrt{q(1-q)} \\
        \sqrt{q(1-q)} & 1-q
    \end{pmatrix},
\end{equation}
while the remaining $R-1$ eigenvalues are equal to $-1/R$. It follows that the trace distance to the mixed stabilizer-code state is
\begin{equation}
\begin{split}
    T_R(q)
    &:=
    \frac{1}{2}
    \left\|
        |\psi\rangle\langle\psi|
        -
        \frac{\Pi_{\mathcal C}}{R}
    \right\|_1
    \\
    &=
    \frac{1}{2}
    \left[
        \frac{R-1}{R}
        +
        \sqrt{
            \left(1+\frac{1}{R}\right)^2
            -
            \frac{4q}{R}
        }
    \right].
\label{eq:mixed_stab_distance}
\end{split}
\end{equation}
We now show that there always exists a pure stabilizer state inside $\mathcal C$ whose distance from $|\psi\rangle$ is no larger than Eq.~\eqref{eq:mixed_stab_distance}. The code $\mathcal C$, of dimension $R=2^m$, can be regarded as the Hilbert space of $m$ logical qubits. A Clifford encoding maps logical stabilizer states onto physical pure stabilizer states contained in $\mathcal C$. Since multiqubit stabilizer states form a complex projective $3$-design
\cite{kueng2015stabilizer3design,zhu2017clifford3design}, defining
\begin{equation}
    x_s = |\langle s|u\rangle|^2
\end{equation}
for the pure stabilizer states $|s\rangle\in\mathcal C$, their second and third moments coincide with the Haar moments,
\begin{equation}
    \mathbb{E}_s[x_s^2]
    =
    \frac{2}{R(R+1)},
    \qquad
    \mathbb{E}_s[x_s^3]
    =
    \frac{6}{R(R+1)(R+2)}.
\end{equation}
Let $x_{\max}=\max_s x_s$. Since
$x_s^3\leq x_{\max}x_s^2$ for every $s$, we obtain
\begin{equation}
    x_{\max}
    \geq
    \frac{\mathbb{E}_s[x_s^3]}
         {\mathbb{E}_s[x_s^2]}
    =
    \frac{3}{R+2}.
\end{equation}
Therefore, there exists at least one pure stabilizer state
$|s\rangle\in\mathcal C$ such that
\begin{equation}
    |\langle s|\psi\rangle|^2
    =
    q\,|\langle s|u\rangle|^2
    \geq
    \frac{3q}{R+2}.
\label{eq:pure_stab_overlap}
\end{equation}
For two pure states the trace distance is
\begin{equation}
    \frac{1}{2}
    \left\|
        |\psi\rangle\langle\psi|
        -
        |s\rangle\langle s|
    \right\|_1
    =
    \sqrt{1-|\langle s|\psi\rangle|^2},
\end{equation}
and hence Eq.~\eqref{eq:pure_stab_overlap} implies
\begin{equation}
    \frac{1}{2}
    \left\|
        |\psi\rangle\langle\psi|
        -
        |s\rangle\langle s|
    \right\|_1
    \leq
    \sqrt{1-\frac{3q}{R+2}}.
\label{eq:pure_stab_distance_bound}
\end{equation}
It remains to compare this expression with the distance to the mixed state. Introducing
\begin{equation}
    \Delta
    =
    \sqrt{(R+1)^2-4Rq},
\end{equation}
one finds from Eq.~\eqref{eq:mixed_stab_distance}
\begin{equation}
\begin{split}
    T_R(q)^2
    -
    \left(1-\frac{3q}{R+2}\right)
    =
    \frac{
        (R-1)(\Delta-1)(R+1-\Delta)
    }{
        2R^2(R+2)
    }.
\label{eq:mixed_vs_pure}
\end{split}
\end{equation}
Since $0\leq q\leq1$, one has
\begin{equation}
    R-1\leq\Delta\leq R+1,
\end{equation}
and therefore the right-hand side of Eq.~\eqref{eq:mixed_vs_pure} is non-negative. Combining this with
Eq.~\eqref{eq:pure_stab_distance_bound} gives
\begin{equation}
    T_R(q)
    \geq
    \frac{1}{2}
    \left\|
        |\psi\rangle\langle\psi|
        -
        |s\rangle\langle s|
    \right\|_1 .
\end{equation}
The case $q=0$ follows trivially, since both distances are equal to one. Thus, for every mixed stabilizer-code state
$\sigma_{\mathcal C}\in\mathrm{STAB}_0^{(N)}$ there exists a pure stabilizer state whose trace distance from the pure target state is no larger. Since pure stabilizer states are themselves contained in $\mathrm{STAB}_0^{(N)}$, we conclude that
\begin{align}
    \min_{\sigma\in\mathrm{STAB}_0^{(N)}}
    \frac{1}{2}
    \left\|
        |\psi\rangle\langle\psi|-\sigma
    \right\|_1
    &= \nonumber \\
    \min_{|s\rangle\in\mathrm{STAB}_{\mathrm{pure}}^{(N)}}
    \frac{1}{2}
    \left\|
        |\psi\rangle\langle\psi|
        -
        |s\rangle\langle s|
    \right\|_1.
\label{eq:mixed_to_pure_stab}
\end{align}
Consequently, for a pure input state the optimization over
$\mathrm{STAB}_0^{(N)}$ can be reduced exactly to an optimization over pure stabilizer states. This allows us to relate the trace-distance problem directly to the optimal stabilizer fidelity.

Applying the previous insight to every locally rotated state gives:
\begin{align}
&\mathcal{M}_{\text{dist}}^{NL}=\nonumber \\
&\underset{U_A,U_B}{\text{min}}  \underset{|s\rangle \in \text{STAB}_{\text{pure}}}{\text{min}} \frac{1}{2} \left\| (U_A \otimes U_B) |\psi \rangle \langle \psi | (U_A^{\dagger} \otimes U_B^{\dagger}) - |s\rangle \langle s |\right\| _1.
\end{align}
For two pure states,
\begin{align}
    \frac{1}{2} \left\| |\phi \rangle \langle \phi |-|s\rangle \langle s | \right\|_1=\sqrt{1-|\langle s| \phi \rangle|^2}.
\end{align}
Therefore
\begin{align}
    \mathcal{M}_{\text{dist}}^{\text{NL}}=\sqrt{1-F_{\text{NL}}(\psi_{AB})},
\end{align}
where
\begin{align}
    F_{\text{NL}}(\psi_{AB}):= \underset{\substack{U_A,U_B\\ |s \rangle \in \text{STAB}_{\text{pure}}}}{\text{max}} |\langle s | U_A \otimes U_B |\psi \rangle|^2
\end{align}
In Ref.~\cite{viscardi2026nonlocal,sierant2026exact} it was proven that the maximal overlap between two pure states with fixed Schmidt coefficients is obtained by aligning their Schmidt vectors in decreasing order. 

For the physical state $|\psi \rangle$, we assume that the unitaries $U_A$ and $U_B$ rotate the states 
$| \phi _i \rangle_A$ and $| \chi_i \rangle _B$, respectively, onto the computational basis:
\begin{align}
    U_A | \phi _i \rangle_A=|i \rangle_A, \; U_B | \chi _i \rangle_B=|i \rangle_B.
\end{align}
This yields:
\begin{align}
    (U_A \otimes U_B) | \psi_{AB} \rangle = \sum_{i=1}^{r_A} \sqrt{\lambda_i} |i \rangle_A | i \rangle_B.
\end{align}
The state is now expressed in a particular stabilizer basis for subsystem $A$ and $B$ stabilized by the Pauli-Z operator. This representation allows for a direct comparison with the pure stabilizer states, which are expressed in the same computational basis.
If the rank of the state is smaller than $\mathrm{min} (d^{n_A},d^{n_B})$, we fix the remaining qubits to a fixed computational basis state, e.g. $|0 \rangle$ without loss of generality. A coherent superposition of $2^r$ computational basis states, with $0 \leq r \leq n_A$, constitutes a pure stabilizer state.
They read:
\begin{align}
    | \sigma_K \rangle = \frac{1}{\sqrt{K}} \sum_{i=1}^K | s_i^A \rangle | s_i^B \rangle, \; K=d^r,
\end{align}
where $|s_i^{A/B} \rangle$ denote computational-basis states and $0\leq r \leq n_A$. The fidelity with these pure stabilizer states is given by:
\begin{align}
    F_K=F(\psi,\sigma_K)=\frac{\Lambda^2_K}{K}, \; \Lambda_K= \sum_{i=1}^{\text{min}(K,r_A)}\sqrt{\lambda_i},
\end{align}
where the $\lambda_i$'s are ordered in decreasing order. The nonlocal magic is then given by:
\begin{align}
M^{\text{NL}}_{\text{dist}} (\psi_{AB}) = \sqrt{1- \underset{K \in \{ 1,d,...,d^{n_A} \}}{\mathrm{max}}F_K}
\end{align}

\section{Fidelity and participation Rényi entropy}\label{app:Fidelity_SRE_relation}
Let us consider the time evolution of a quantum many-body system starting from an initial product state $\ket{\psi(0)}\!=\!\ket{\vb*{\alpha}_{0}}$ where $\ket{\vb*{\alpha}_{0}}$ is a specific state of the computational basis.
Then the fidelity $\mathcal{F}(t)\!=\!\abs{\braket{\psi(0)|\psi(t)}}^{2}$ reads
\begin{equation}
    \mathcal{F}(t)=\abs{C_{\vb*{\alpha}_{0}}(t)}^{2}\equiv p_{0}(t).
\end{equation}
Assuming the time-evolved state to be normalized and defining probabilities as $p_{i}(t)\!=\!\abs{C_{\vb*{\alpha}_{i}}(t)}^{2}$ with $\sum_{i}p_{i}\!=\!1$ and $p_{0}\!=\!p_{i_{0}}$, we observe that
\begin{equation}
   \mathrm{IPR}_{2}(t) = \sum_{i}p_{i}(t)^{2}= p_{0}(t)^{2}+ \sum_{i\neq i_{0}}p_{i}(t)^{2},
\end{equation}
where $\mathrm{IPR}_{2}$ is the inverse participation ratio~\cite{Liu2025QuantumAlgorithmsInverse}.
For a fixed $p_{0}$, the rest of $\mathrm{IPR}_{2}$ is bounded as follows: the minimum is when the remaining weight is uniformly spread over the other $d^{N}-1$ states of the computational basis; the maximum is when the remaining weight sits on a single other basis state. 
Namely:
\begin{equation}
    \frac{(1-p_{0})^{2}}{d^{N}-1}\leq 
    \sum_{i\neq i_{0}}p_{i}^{2}
    \leq (1-p_{0})^{2},
\end{equation}
which implies that
\begin{equation}
\begin{split}
    \mathrm{IPR}_{2}&
    \leq p_{0}^{2}+ (1-p_{0})^{2},\\
    \PE_{2}&\geq-\log_{2}(p_{0}^{2}+ (1-p_{0})^{2}),\\
    \PE_{2}(t)&\geq
    -\log_{2}\qty(\mathcal{F}(t)^{2}+ (1-\mathcal{F}(t))^{2}),
\end{split}
\label{eq_bound1_fidelity_PE}
\end{equation}
and correspondingly
\begin{equation}
\begin{split}
    \mathrm{IPR}_{2}&\!\geq\!p_{0}^{2}+
    \frac{(1-p_{0})^{2}}{d^{N}-1}\geq p_{0}^{2},\\
    \PE_{2}&\leq\!-\!\log_{2}\qty(p_{0}^{2}+
    \frac{(1-p_{0})^{2}}{d^{N}-1})\leq -\log_{2}(p_{0}^{2}),\\
    \PE_{2}(t)&\!\leq\!
    -\!\log_{2}\qty[\mathcal{F}(t)^{2}\!+\!
    \frac{(1\!-\!\mathcal{F}(t))^{2}}{d^{N}\!-\!1}]\!\leq \!-\!2\log_{2}(\mathcal{F}(t)).
\end{split}
\label{eq_bound2_fidelity_PE}
\end{equation}
The upper bound in \cref{eq_bound2_fidelity_PE} decreases monotonically as $\mathcal{F}(t)\to1$ and forces $\PE_{2}(t)\to0$: this is relevant to the scarring revivals of Fig.~\ref{fig:PXP_dynamics}, where a growing weight on the initial basis state maximizes $\sum_{i}p^{2}_{i}$ and suppresses $\PE_{2}$. 
We stress that the lower bound in \cref{eq_bound1_fidelity_PE} is not monotonic in $\mathcal{F}$: it attains its maximum $\log_{2}2$ at $\mathcal{F}=1/2$ and vanishes at both $\mathcal{F}\to0$ and $\mathcal{F}\to1$, so it constrains $\PE_{2}$ only from below and carries no information about the revivals.

An upper bound on the SRE in terms of the participation entropy was derived in Ref.~\cite{Tirrito2025UniversalSpreadingNonstabilizerness}:
\begin{align}\label{eq_SRE_PE2}
    \mathcal{M}_{\text{SRE}}(t)\leq 2 \text{PE}_2(t).
\end{align}
Combining \cref{eq_bound2_fidelity_PE} with the PE bound on SRE in \cref{eq_SRE_PE2} finally yields the fidelity bound on the global magic quoted in the main text,
\begin{equation}
    \mathcal{M}_{\text{SRE}}(t)\;\leq\;2\,\PE_{2}(t)\;\leq\;-4\log_{2}\mathcal{F}(t),
    \label{eq_fidelity_SRE_bound}
\end{equation}
which is the chain underlying \cref{eq_M2_fidelity}. 

\section{Ergodicity of the models}\label{app:ergodicity}
In this paper, we look only at quantum-chaotic models, \idest{}, models whose energy spectrum behaves similarly to that of a generic member of the Gaussian orthogonal ensemble (GOE). 
As a reference point, we use the restricted gap ratio, defined as
\begin{equation}
    r_n=\min(\tilde{r}_n,1/\tilde{r}_n),
    \label{eq:def_restricted_gap_ratio}
\end{equation}
where $\tilde{r}_n\!=\!s_n/s_{n-1}$ and $s$ is the difference between two adjacent eigenvalues, sorted from smallest to largest. 
We compare the distribution of $r_n$ and its mean $\langle r \rangle$ with the GOE prediction $\langle r_{\mathrm{GOE}} \rangle\!=\!0.535898$ and with the prediction of an uncorrelated (Poisson) spectrum $\langle r_{\mathrm{P}} \rangle\!=\!0.386294$.

\begin{figure*}
\includegraphics[width=1.0\textwidth]{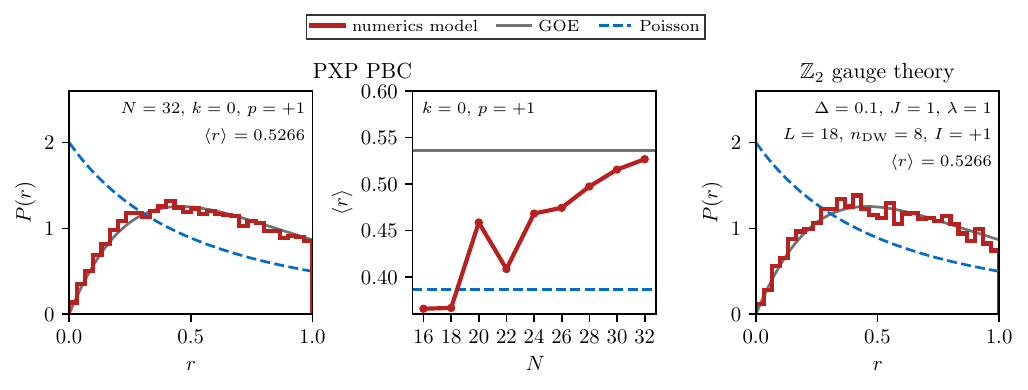}
    \caption{The distribution of the restricted gap ratio of the PXP model of size $N=32$, momentum sector $k=0$ and parity sector $p=+1$, compared with the GOE and Poisson predictions (left). The mean restricted gap ratio of the same symmetry sector with respect to the system size (middle). The distribution of the restricted gap ratio of the $\mathbb{Z}_2$ gauge theory of size $N=18$ and parity sector $p=+1$ compared to the GOE and Poisson prediction (right).}
    \label{fig_PXP_Iadecola_r_value}
\end{figure*}

These quantum-chaotic spectral statistics are consistent with the applicability of ETH and with thermalization of generic high-temperature states. For the PXP model with $N\!=\!32$ in the $k\!=\!0$ momentum sector and $p\!=\!+\!1$ parity sector, we find a mean restricted gap ratio of $\langle r_{\mathrm{PXP}} \rangle\!=\!0.5266$, with the distribution closely following the GOE prediction, as shown in the left panel of \cref{fig_PXP_Iadecola_r_value}.
In particular, the middle plot shows that the PXP model only becomes truly quantum chaotic at relatively large system sizes. 
In our analysis, we are unable to compute the more expensive observables, such as the SRE, for $N=32$. Nevertheless, already at $N=18$, most of the eigenstates exhibit entanglement scaling consistent with the ergodic volume law, whereas the special scar states display logarithmic scaling.

In contrast, the $\mathbb{Z}_2$ gauge theory has much stronger signatures of ergodicity at smaller $N$, with a restricted gap ratio of $\langle r \rangle_{\mathbb{Z}_2}=0.5266$ already at $N=18$ in the parity sector $p=+1$. The corresponding restricted gap ratio distribution is shown in the right panel of \cref{fig_PXP_Iadecola_r_value}.

\section{SRE for exact $\mathbb{Z}_2$ scars}\label{app:SRE_Z2_scars}
We consider the exact scar state $|S_n \rangle$ introduced in the main text for the Hamiltonian defined in Eq.~\eqref{eq_Z2_hamiltonian}. The scar states can be written as:
\begin{align}
    |\mathcal{S}_n\rangle = \frac{1}{\mathcal{N}} 
    \sum_{\substack{\text{configs. with }\\ n \text{ excitations,} \\ 
    \text{no two adjacent}}} 
    (-1)^{\sum_{p} i_p} \;|0 \cdots \underset{i_1}{1}  \cdots \underset{i_n}{1} \cdots 0\rangle,
\end{align}
where $i_p$ is the position of the $p$-th excitation,
and $\mathcal{N} = \sqrt{\binom{L-n+1}{n}}=\sqrt{\binom{N-n-1}{n}}$. We set $L=N-2$. Under open boundary conditions, both boundary spins are fixed $x_1=0$ and $n_N=0$. The bulk sites are $2,3,...,N-1$. As we will work in the following mainly with the bulk sites, let us relabel them for simplicity: $x=(x_1,x_2,...,x_L) \in \{0,1\}^L$. The set of valid bulk configurations is:
\begin{align}
    \mathcal{I}_{L,n} &:= \{ x=(x_1,...,x_L) \in \{0,1\}^L  \Big| \sum_j^Lx_j
    =n \; \nonumber\\
    &\text{and} \; x_j \cdot x_{j+1}=0 \; \forall j \in \{ 1,...,L-1 \} \} .
\end{align}
The scar state is characterized via two constraints, a fixed excitation number and no two excitations next to each other. The number of such configurations is:
\begin{align}
    |\mathcal I_{L,n}| = \binom{L-n+1}{n} = \binom{N-n-1}{n}, 
\end{align}
The scar state can equivalently be expressed as:
\begin{align}
    |S_n \rangle := \frac{1}{\sqrt{N_{N,n}}} \sum_{x \in \mathcal{I}_{L,n}} (-1)^{\sum^L_{j=1} jx_j} |x\rangle.
\end{align}
Here the phase factor $(-1)^{\sum^L_{j=1} jx_j}$ is simply $(-1)^{N_{\text{odd}}}$, where $N_{\text{odd}}$ denotes the number of excitations on the odd sites of the chain.

The second Stabilizer-Rényi entropy is $\mathcal{M}_{\text{SRE}}(\psi)=-\text{log}_2 \Xi_2(\psi)$ with:
\begin{align}
    \Xi_2 (\psi):= 2^{-L} \sum_{p,q \in \mathbb{F}^L_2} |\langle \psi | X^p Z^q | \psi \rangle|^4,
\end{align}
where $X^p=\bigotimes^L_{j=1}X^{p_j}$ and $Z^q=\bigotimes^L_{j=1} Z^{q_j}$. The staggered phase can be written by the Clifford unitary:
\begin{align}
    U_{\text{stag}}=\bigotimes^L_{\substack{j=1\\j \text{ odd}}}Z_j, \; U_{\text{stag}} |x \rangle = (-1)^{N_{\text{odd}(x)}}|x\rangle.
\end{align}
Therefore, $|S_n \rangle = U_{\text{stag}} |\Phi_n \rangle$. As $U_{\text{stag}}$ is Clifford, it leaves the SRE invariant $\mathcal{M}_{\text{SRE}}(S_n)=\mathcal{M}_{\text{SRE}}(\Phi_n)$. A Pauli matrix element can be written:
\begin{equation}
A_{\bm{p},\bm{q}} = \frac{1}{\mathcal N_{N,n}} \sum_{\bm{x}\in\mathcal I_{L,n}} \chi_{L,n}(\bm{x}\oplus\bm{p}) (-1)^{\bm{q}\cdot\bm{x}}, \label{eq:pauli-expectation} 
\end{equation} 
where $\chi_{L,n}$ denotes the indicator function of $\mathcal I_{L,n}$: 
\begin{equation} 
\chi_{L,n}(\bm{x}) = \begin{cases} 1, & \bm{x}\in\mathcal I_{L,n}, \\ 0, & \bm{x}\notin\mathcal I_{L,n}. 
\end{cases} 
\end{equation}
Using the Walsh-Fourier identity gives:
\begin{align}\label{eq:K-definition} 
K_{L,n} ={}& \sum_{\bm{p}\in \mathbb{F}^L_2} \sum_{\bm{x}_1,\ldots,\bm{x}_4\in\mathcal I_{L,n}} \delta_{ \bm{x}_1 \oplus\bm{x}_2 \oplus\bm{x}_3 \oplus\bm{x}_4, \bm{0} } \nonumber\\ &\times \prod_{a=1}^{4} \chi_{L,n}(\bm{x}_a\oplus\bm{p}). 
\end{align}
In the end we have the formula:
\begin{equation} 
\mathcal{M}_{\text{SRE}}(S_n) = 4\log_2\mathcal N_{N,n} - \log_2K_{L,n} . 
\label{eq:M2-exact} 
\end{equation} 
It remains to calculate the integer $K_{L,n}$, which can be calculated exactly. This exact enumeration yields the blue pentagons in the right panel of Fig.~\ref{fig:Nl_magic_static}. To carry out the enumeration, we first introduce a local description for each bulk site:
\begin{align}
    s^{(j)}= \left( x_1^{(j)},x_2^{(j)},x_3^{(j)},x_4^{(j)},p^{(j)} \in \{ 0,1 \} \right).
\end{align}
This local description consists of $4$ replica sites and the shift string $p^{(j)}$. At each site, the XOR constraint imposes $x_4^{(j)}=x_1^{(j)} \oplus x_2^{(j)} \oplus x_3^{(j)}$ resulting in $16$ distinct local states. The shifted replicas variables are given by $y_a^{(j)}=x_a^{(j)} \oplus p^{(j)}$. Two consecutive local states $s$ and $s'$ at sites $j$ and $j+1$ are compatible if
\begin{align}
    x_a x_a'=0 \; \; y_ay_a'=0 \; \text{for each }a \in \{ 1,2,3,4 \}.
\end{align}
We now enumerate all allowed transitions between local states, from the left boundary to the right while enforcing both constraints.
With this method, we can calculate $\mathcal{M}_{\text{SRE}}$ exactly up to $N=56$.

We next derive the asymptotic large-$N$ formula for the $\mathcal{M}_{\text{SRE}}$. To take the large-$N$ limit, we use the approach of fugacities and a saddle point approximation to impose the constraint of a certain fixed magnon number $n$. 
We introduce eight fugacities $z_a,w_a \; (a=1,2,3,4)$ for the $x$ and $y$ replicas. As all replicas are treated equally, we set $z_a=w_a=t$. We first define the corresponding allowed-transition matrix:
\begin{align}
    \mathcal{A}(s,s'):= \prod_{a=1}^{4} (a-x_ax_a') \cdot \prod_{a=1}^{4} (a-y_ay_a')=1.
\end{align}
The fugacity-weighted transfer matrix reads:
\begin{align*}
    T_{s,s'}(z,w)=\mathcal{A}(s,s') \prod_{a=1}^4 t^{x_a'+y_a'}. 
\end{align*}
The generating function is:
\begin{align}
    \mathcal{Z}_L(z,w)=\langle v_L | T(z,w)^L |v_R \rangle,
\end{align}
where $K_{L,n}$ is the coefficient of the term proportional to $t^{8n}$ in $\mathcal{Z}_L(z,w)$. The fugacity weight depends only on $k=x_1+x_2+x_3+x_4$ and $p$. Imposing the XOR constraint leaves $6$ valid states. For $m_{k,p}$: $m_{0,p}=1$, $m_{2,p}=6$ and $m_{4,1}$. The fugacity factor reads:
\begin{align}
w(s')=\prod^4_{a=1}t^{x'_a+y'_a}=t^{\sum_{a=1}^4(x'_a+y'_a)}.
\end{align}
For $p'=0$, one has $y'_a=x_a$, thus $\sum_a(x'_a+y_a')=2k$. The weight factor is $w(s')=t^{2k}$. For $p'=1$, we get $y'_a=1-x'_a$, and therefore $\sum_a(x'_a+y_a')=4$ and consequently $w(s')=t^4$. Now one can rewrite the transfer matrix with the matrix entries labeled by $(k,p)$ and $(k',p')$. Equivalently, this can be written as $[T]_{(k,p),(k',p')}=w(k',p') \cdot \# \{\text{states s' in sector (k',p') compatible with s}\}$. This gives  a $6 \times 6$ $T$ matrix. The generating function reads:
\begin{align}
    Z_{L}(t)&=\langle v_L |T^L | v_R \rangle=\sum_{i=1}^6 \lambda^L_i \langle v_L | \phi^R_i \rangle \langle \phi^L_i | v_R \rangle \nonumber\\
    &=\sum_{i=1}^6 c_i(t) \lambda_i(t)^L.
\end{align}
Recall that our goal is to extract the coefficient $K_{L,n}=[t^{8n}]\mathcal{Z}_L(t)$. In the large-$L$ limit, the generating function is dominated by the largest eigenvalue $\lambda_{*}$ of the transfer matrix, i.e., the Perron--Frobenius eigenvalue. This can be formally written via a contour integral:
\begin{align}
    K_{L,n}&=\oint \frac{dt}{2 \pi i t^{8n+1} }\sum_{i=1}^6 c_i(t) \lambda_i(t)^L \nonumber\\
    &=\sum_{i=1}^6\oint \frac{dt}{2 \pi i t^{8n+1} } c_i(t) \lambda_i(t)^L \nonumber\\
    &=\sum_{i=1}^6\oint \frac{dt}{2 \pi i t } c_i(t)e^{L \text{ln}(\lambda_i(t))-8n \text{ln}(t)}.
\end{align}
For large $L$,  each integral is dominated by the corresponding saddle point of $h_i(t)=L \text{ln}(\lambda_i(t))-8n$. Expanding the exponent to second order around the saddle point, the integral can then be evaluated within the saddle-point approximation. The saddle-point condition, $h'_i(t)=0$ yields $t \frac{d \text{ln}(\lambda_i))}{dt}=\frac{8n}{L}$. 
Writing $n=N/k$ and using $L=N-2$, this condition reduced in the large-$L$ limit to $t \frac{d \text{ln}(\lambda_i))}{dt}=\frac{8}{k}$. 
This can also be written as $u \frac{1}{\lambda_i} \frac{d \lambda_i}{du}=\frac{2}{k}$ with $u=t^4$. The characteristic polynomial of the transfer matrix $T$ is given by:
\begin{align}
    (\lambda-u)^2(\lambda+u) [\lambda^3 - (1+u)\lambda^2-(u^2+13u)\lambda+u^3],
\end{align}
with $u=t^4$.
Two solutions are $\lambda=u$ (double) and $\lambda=-u$. Let us define the last part as:
\begin{align}\label{eq:F_characteristic}
   F(\lambda,u)=\lambda^3 - (1+u)\lambda^2-(u^2+13u)\lambda+u^3.
\end{align}
For the remaining $3$ eigenvalues, this term must vanish. We can differentiate $F(\lambda(u),u)$ with respect to $u$
\begin{align}
    \frac{\partial F}{\partial \lambda} \frac{d \lambda_i}{d u} + \frac{\partial F}{\partial u}=0.
\end{align}
Substituting the saddle point condition then yields:
\begin{align}
    &\frac{\partial F}{\partial \lambda} \frac{ \lambda_i2}{k} + \frac{\partial F}{\partial u}=0\\
    &=2u\frac{\partial F}{\partial u} + \lambda_i \frac{\partial F}{\partial \lambda}.
\end{align}
We are left with two polynomial equations, namely the equation above and $F(\lambda,u)=0$. Both have to hold at the saddle point. Through elimination, one gets:
\begin{align}
    u^5(448u^3-640u^2+868u-507)=0.
\end{align}
We get also complex solutions for $t$. Since $K_{L,n}$ is real, we restrict our analysis to the real saddle point. In the large-$L$ limit, the contribution from the largest real eigenvalue dominates the integral, while contributions from the remaining eigenvalues are exponentially suppressed.

Expanding the exponent in the contour integral about the saddle point gives: 
\begin{align}
    K_{L,n}&=\oint \frac{dt}{2 \pi i t^{8n+1} }\sum_{i=1}^6 c_i(t) \lambda_i(t)^L \nonumber \\
    &=\sum_{i=1}^6\frac{c_i(u_i*)}{\sqrt{2 \pi L \sigma^2}} e^{L F_i} +O(L^{-1}),
\end{align}
with $F_i=\text{ln}(\lambda_i(u_i*))-2n/L \text{ln}(u_i*)$. The largest eigenvalue gives the dominant contribution. Let us denote this dominant value by $F*=\text{max}(F_i)$:
\begin{align}
    K_{L,n}&=\frac{c_i(u_i*)}{\sqrt{2 \pi L \sigma*^2}}e^{LF*} \left[ 1+\sum_{i \neq*} \frac{c_i(u_i^*)}{c_*(u_*)}\sqrt{\frac{\sigma^2_*}{\sigma_i^2}} e^{-K(F_*-F_i} \right] \nonumber \\
    &+ O(L^{-1}).
\end{align}
The real positive solution is:
\begin{align}
    u_*\approx0.7897448219.
\end{align}
Substituting this value into Eq.~\eqref{eq:F_characteristic} yields:
\begin{align}
    \lambda_*\approx 4.297306133.
\end{align}
For $n=N/4$, the large-$N$ expression becomes
\begin{align}
    \mathcal{M}_{\text{SRE}}(S_{N/4})\approx 0.4812N+2 \text{log}_2N+C,
\end{align}
where $C$ is some constant.

\section{SRE for exact PXP scars}\label{app:SRE_PXP_scar_exact}
For the PXP model with open boundary conditions, four exact scar eigenstates admit a finite-bond-dimension matrix-product-state (MPS) representation~\cite{Smith2025NonstabilizernessKineticallyConstrained, Lin2019ExactQuantumManyBody}:
\begin{align}
    \ket{\Gamma_{\alpha,\beta}} &= \sum_{ \{ \sigma \} } v^T_{\alpha} A^{\sigma_1 \sigma_2} A^{\sigma_3 \sigma_4} 
    \cdots A^{\sigma_{N-1} \sigma_N} v_{\beta} \ket{\sigma_1\cdots\sigma_N},
\end{align}
where $\alpha,\beta\!\in\!\qty{1,2}$, the boundary vectors are given by $v_1\!=\!(1,1)^T$, $v_2\!=\!(1,-1)^T$ and the matrices $A^{\sigma_i \sigma_{i+1}}$ are defined as follows:
\begin{align}
    A^{00}\!=\!\begin{pmatrix}
    0 & -1 \\
    1 & 0 
\end{pmatrix}\!,
A^{01}\!=\!\begin{pmatrix}
    \sqrt{2} & 0 \\
    0 & 0 
\end{pmatrix}\!,
A^{10}\!=\!\begin{pmatrix}
    0 & 0 \\
    0 & -\sqrt{2}
\end{pmatrix}.
\end{align}
The fourth matrix vanishes, $A^{11}=0$, thereby imposing the Rydberg-blockade constraint in the MPS~\cite{Lin2019ExactQuantumManyBody}.
The four scar states have energies $E=0$ for $\ket{\Gamma_{11}}$ and $\ket{\Gamma_{22}}$, and $E=\pm\sqrt{2}$ for $\ket{\Gamma_{12}}$ and $\ket{\Gamma_{21}}$. The latter two are Hamiltonian eigenstates with definite symmetry quantum numbers, whereas $\ket{\Gamma_{11}}$ and $\ket{\Gamma_{22}}$ do not have definite parity. Definite parity states are recovered by taking the symmetric and antisymmetric combinations. 
The normalization of the states is $Z_{\alpha \beta}=\langle \Gamma_{\alpha \beta}|\Gamma_{\alpha \beta} \rangle$. The state is identical on every two-site unit cell framed by the two boundary vectors. This can be used to write the overlap as $Z_{\alpha \beta}=E_{\alpha}^{(1)^T} T_{1}^{N/2}E_{\beta}^{(T)}$, where $T_1=\sum_{\sigma \in \{ 00,01,10\}} A_{\sigma} \otimes A_{\sigma}$ is the transfer-matrix and $E_\alpha=v_\alpha \otimes v_\alpha$ is composed of the boundary vectors. The eigenvalues of $T_1$ are $\lambda \in \{ 3,1,1,-1 \}$.

To evaluate the expectation value of a Pauli string in the scar state, appearing in the quantity $|\langle \psi | P | \psi \rangle|^4$, we exploit the two-site unit-cell structure and factorize the Pauli string accordingly. We denote the Pauli operator acting on a single unit cell by $Q \in \{I,X,Y,Z \}^{\otimes 2}$. This leads to the corresponding dressed transfer matrix:
\begin{align}
    T_{\text{SRE}}&=\sum_Q D(Q)^{\otimes 4}, \\
     D(Q)&=\sum_{\sigma \sigma' \in \{00,01,10\}} Q_{\sigma \sigma'}A^{\sigma} \otimes A^{\sigma'}. \nonumber
\end{align}
Finally we get for the SRE:
\begin{align}
    \mathcal{M}_{\text{SRE}}(\Gamma_{\alpha \beta})&=- \log_2 \qty( \sum_a p_a^2 ) - \log_2 (D),\\
    \sum_a p^2_a &=  \frac{ E_{\alpha}^{(4)^T} T_{\mathrm{SRE}}^{N/2} E_{\beta}^{(4)}}{D^2Z_{\alpha \beta}^4}. \nonumber
\end{align}
We now evaluate this expression explicitly for the scar states $\Gamma_{12}$ and $\Gamma_{21}$. Because we are interested in the asymptotic scaling with $N$, we invoke the Perron-Frobenius theorem, according to which the contribution from the largest eigenvalue dominates in the large-$N$ limit. The normalization gives us: $Z_{12}=Z_{21}=\lambda_{\text{max}} (E_1^{(1)^T} R_{\text{max}}  L_{\text{max}}^T E_2^{(1)})=3^{N/2}2$, where $R_{\text{max}}$ and $L_{\text{max}}$ are the right and left eigenvectors associated with the largest eigenvalue of $T_1$. For the dressed transfer matrix $T_{\text{SRE}}$, the largest eigenvalue is $\lambda_{\text{SRE}, \text{max}}=152.6640$. The corresponding contribution to $|\langle \psi | P | \psi \rangle|^4$ is given asymptotically by $\lambda_{\text{SRE}, \text{max}} (E_\alpha^{(4)^{\dagger}}R_{\text{SRE}, \text{max}})(L_{\text{SRE}, \text{max}}^{\dagger}E_\beta^{(4)})$, where $E_\alpha^{(4)}=v_\alpha^{\otimes 8}$. For both $\Gamma_{12}$ and $\Gamma_{21}$, the boundary-vector contribution evaluates to $13.2915$, yielding $|\langle \psi | P | \psi \rangle|^4=152. 6640^{N/2}13.2915$.
 \begin{align}
     \mathcal{M}_{\text{SRE}}(\Gamma_{12})&=\mathcal{M}_{\text{SRE}}(\Gamma_{21}) \nonumber\\
     &=-\text{log}_2 \left( \frac{\sum_P|\langle \Gamma_{12} | P | \Gamma_{12} \rangle|^4}{D^2Z^4_{12}} \right) - \text{log}_2(D),\nonumber\\
     \Rightarrow&-\text{log}_2 \left( \frac{152.6640^{N/2}13.2915}{D^2(3^{N/2}2)^4} \right) -\text{log}_2(D)\nonumber\\
     &=s_2N+c, \; s_2=0.5428, \; c=0.2676.
 \end{align}

\newpage
\bibliography{bibliography_paper}
\end{document}